\documentclass[12pt]{iopart}
\usepackage{iopams,bbm}
\usepackage{epsfig}
\begin{document} 
\newtheorem{prop}{Proposition}
\newtheorem{theo}{Theorem}

\title{Quantum walks on Cayley graphs}
\author{O Lopez Acevedo\dag\ddag and T Gobron\dag}
\address{\dag Laboratoire de Physique Th\'eorique et Mod\'elisation,
Universit\'e de Cergy-Pontoise, 2 Avenue Adolphe Chauvin 95302 Cergy
Pontoise Cedex, France}
\address{\ddag Institut f\"ur Mathematik und Informatik,
  Ernst-Moritz-Arndt-Universit\"at, Friedrich-Ludwig-Jahn Str. 15a, 17487
  Greifswald, Germany} 
\eads{\mailto{lopez@ptm.u-cergy.fr}}

\begin{abstract}
We address the problem of the construction of quantum walks on Cayley graphs.
Our main motivation is the relationship between quantum algorithms and quantum
walks. 
In particular, we discuss the choice of the dimension of the local Hilbert
space and consider various classes of graphs on which the structure of quantum 
walks may differ. 
We completely characterise quantum walks on free groups
and present partial results on more general cases. Some examples are given
including a family of quantum walks on the hypercube involving a Clifford 
Algebra.  
\end{abstract}

\pacno{03.67.Lx}

\maketitle

\section{Introduction}
Recently many effort has been devoted to the construction of new quantum
algorithms. In particular a question which has arisen is whether the
known algorithms fully exploit the possibilities of quantum mechanics,
or if there could exist more efficient ones.
A search for new ideas in this direction has been at the origin of a renewed
study of quantum walks models \cite{ambainis_algo_qrw}, and a few
results have already been obtained, showing that these are definitely relevant
in this context.\\
The first general characterisation of walks over graphs was presented in
\cite{aakv}. A possible construction for a walk operator is given, based on
its classical equivalent, and some quantities relevant in the context
of quantum algorithms are defined and computed.
One of the principal results states that for bounded degree graphs the mixing
time (defined also in the same work) is at most quadratically faster than the
mixing time of the simple classical random walk on the same graph. Even if
this general result is not so encouraging, some particular graphs have
been shown to have properties intrinsically different from their classical
equivalents. In particular, a symmetric quantum walk may
get across an hypercube in a time linear with the dimension, while its
classical counterpart would take an exponentially larger time.\\
In algorithmic applications, quantum walks have also shown interesting
properties. The first important achievement has been the setting of the
quantum search algorithm in the form of a quantum walk over an hypercube
\cite{Shenvi}.
Some other similar quantum search algorithms were constructed
after this. In one of them \cite{akr04}, the choice of the coin operator
was revealed to be of crucial importance, since different operators may achieve
different speed-ups (or no speed-up at all) without obvious reasons. A natural
question which arises from this problem is whether there exist quantum walks
different to than those defined in \cite{aakv} and if so, to what extent they
could be the source of interesting new properties and algorithmic
applications. Another problem lies in the dimension of the internal space: it
is always possible to enlarge it, and in \cite{brun_ambainis}, it was shown
that in an extremal case, the variance of the one dimensional walk recovers the
classical behaviour. In a similar direction in \cite{WD03} and \cite{WD04} the
authors have considered the evolution of a quantum particle governed by a
quantum multi-baker map which can be settled as a quantum walk on a line with a
multidimensional internal space, the classical limit is also recovereded
enlarging the dimension of the internal space. At the opposite, an interesting
and still open question is whether there exist quantum walks with local spaces
of dimension smaller than that taken in the standard definition.
In the context of quantum cellular automata, it is shown in \cite{meyer2} that
for the simple lattice in $d$ dimensions there is no nontrivial walk with
an internal space of dimension one, also known as the No-go theorem.\\
In this article we make a step in the direction of determining all
possible quantum walks for general graphs and characterising their
structures. Starting from a general definition of a quantum walk we deduce
necessary and sufficient conditions on the coin operators for the evolution to
be unitary (section 2). The next section contains a discussion on the
solutions of these equations (section 3). In particular, we characterise
all possible walks on the Cayley graph of a free group. In the case of
abelian groups, the situation is somewhat
more complicated, and after a general discussion we present particular
solutions. We construct quantum walks over the two dimensional and three
dimensional simple lattice with an internal space of dimension smaller than
what was previously known and a generalisation to arbitrary dimensions. We
also consider the hypercube as a Cayley graph on which we construct a quantum
walk where the coin operators are related to elements of the Clifford
algebra. Finally, we propose a possible generalisation of a quantum  walk
where we depart from the image of a particle moving on a lattice and which
could be of interest in the context of quantum algorithms (section 4).
\section{Model and unitary relations}
A quantum algorithm is a sequence of transformations on a state of a
quantum system. The quantum system is described by a tensor product of two
dimensional complex Hilbert spaces. There is a preferred basis of the
elementary space where vectors are labelled with the integers zero and one in
correspondence to classical bits. Then a basis vector of the entire system is
$|x_0 \rangle \otimes \dots \otimes | x_n \rangle$ where $x_i \in \{0,1\}$ and
in this way it is possible to associate to each base vector an integer whose
binary decomposition coincides with the n-tuple $(x_0, \dots ,x_n)$. The total
operator is the product of elementary operators. A presentation of possible
sets as well as a demonstration of the universality of these sets may be found
in \cite{barenco}.\\
A quantum walk is a model for the evolution of a particle over a
graph. Many of the choices made in building  the model may be explained by the
aim  of studying them as quantum algorithms. Let $G$ be a directed graph with
vertex set $X$ and edge set $E$ such that $G=(X,E)$. Let $\mathcal H$ be the
Hilbert space defined by $\mathcal H= \mathcal H_I \otimes \mathcal H_G$.
The space $\mathcal H_G=\ell^2(X)$ describes the position of the
particle over the graph and the space $\mathcal H_I= \mathbb{C}^d$ describes
some internal degrees of the particle.Let $\{|x \rangle\}_{x \in X}$ be a base
of $H_G$ and $\{ |1\rangle ,\dots, |d\rangle\}$ a base of $\mathcal H_I$.\\
The evolution equation is:
\begin{equation}
|\psi_{t+1}\rangle=W |\psi_{t} \rangle
\end{equation}
where $W$ is a discrete time evolution operator defined as
\begin{equation}\label{eo}
W=\sum_{x\in X}\sum_{z \in E_x} M_{x,z} \otimes T_{x \rightarrow z}
\end{equation}
where $E_x$ denotes the set of neighbouring sites of $x$ and $T_{x \rightarrow
  z}$ translates the particle from $x$ to $z$. $T_{x \rightarrow z}$ is
  defined by 
\begin{equation}
\langle x' | T_{x \rightarrow z} \vert \psi\rangle = \langle x'|z \rangle
\langle x | \psi \rangle
\end{equation}
$M_{x,z}:H_I \to
H_I$ are maps modifying the internal space at the same time as the translation
from vertex $x$ to vertex $z$ is applied. Suppose $|\psi_{t}\rangle=\vert i
\rangle \otimes |z \rangle$. Then after one time step the probability of
finding the particle in at vertex $y$, a neighbour of $z$, will depend on the
previous internal state: 
\begin{equation}
P(y)= \sum_{j=1}^d | \, \langle j | M_{z,y} |i \rangle |^2
\end{equation}
One image commonly used to describe the local evolution is that of a coin
attached to each vertex and flipped to decide which neighbour the
particle will jump to (see for instance \cite{aakv}) and accordingly the local
map $M_{x,y}$ is termed the ``coin operator''. Here we follow this usage
though our model is more general than the image: in fact,
it is important to note that originally the internal state was identified to
the set of possible outcomes of the coin flip, or equivalently to the set of
neighbours, so that the dimension of the internal space at a given vertex was
necessarily equal to the number of outgoing edges. Here we have not considered
this identification.\\ 
Unitarity of $W$ is satisfied if and only if:
\begin{eqnarray}
W^\dagger W=\mathbbm{1} \Leftrightarrow \sum_{z \in E_x \cap E_{x'}}
M_{x,z}^\dagger M_{x',z}=\delta_{x,x'} \mathbbm{1}_{H_I} \label{co1} \\
W W^\dagger=\mathbbm{1} \Leftrightarrow \sum_{z \in E_x  \cap E_{x'}} M_{z,x}
M_{z,x'}^\dagger=\delta_{x,x'} \mathbbm{1}_{H_I} \label{co2}
\end{eqnarray}
$\forall  x,x'$. When  $x \not = x'$, in order to have a non trivial equation,
$x$ and $x'$ must be second neighbours and the number of terms in the sum is
related to the number of closed paths of length 4 with alternating
orientation.\\ 
In the example on the figure 1, one condition equation of the form \eref{co1}
with three terms is associated with the pair of second neighbours $x$ and
$x'$ : 
\begin{equation}
M_{x,z_1}^\dagger M_{x',z_1}+M_{x,z_2}^\dagger M_{x',z_2}+M_{x,z_3}^\dagger
M_{x',z_3}=0 
\end{equation}
\begin{figure}
\begin{center}
\epsfbox{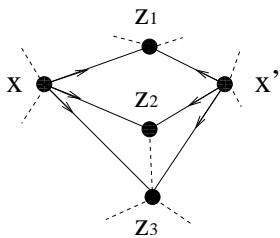}
\end{center}
\caption{A pair of second neighbours and all paths of length two
  between them}
\end{figure}

\section{Quantum walks on Cayley Graphs}
We will restrict our study from now on to quantum walks
on Cayley graphs. We first recall their definitions and main properties.
We follow the presentation given in \cite{white}.
Given a group $\Gamma$ one considers a set $\Delta$ of elements in $\Gamma$
such that $\Delta$ is a generating set for $\Gamma$.
The Cayley graph $C_\Delta(\Gamma)=(X,E)$ is defined as the oriented graph
with 
\begin{eqnarray}
X\equiv X(C_\Delta(\Gamma))=\Gamma \label{xcg}\\
E\equiv E(C_\Delta(\Gamma))=\{(x,x\delta)_{\delta} \vert x\in \Gamma ,\delta
\in \Delta \} \label{ecg}
\end{eqnarray}
When associating a colour to each element of the generating family, the
definition of $C_\Delta(\Gamma)$ makes it a coloured directed graph. In
addition a Cayley colour graph is vertex transitive, so that each site is
equivalent. Thus we consider internal operators which depend only on the
edge colour and direction of the edge $(x,y)$ (i.e. only on the generator
$\delta= x^{-1} y$) and not on the starting vertex $x$:
\begin{eqnarray}
M_{x,y}= M_{x^{-1} y} \hbox{ for all } (x,y) \in E
\end{eqnarray}
Thus the evolution operator $W$ on $\mathcal H$ is
\begin{eqnarray}
\label{wcg}
W = \sum_{\delta\in\Delta} M_\delta \otimes T_\delta
\end{eqnarray}
where $T_\delta$ is the shift in the direction $\delta$ and is defined
for all vertices by the group operation
\begin{equation}
T_\delta = \sum_{x \in X} T_{x \rightarrow x \delta}
\end{equation}
The problem is thus reduced to a local one on $\mathcal H_I$
and the unitarity conditions \eref{co1} and \eref{co2} now read:
\begin{eqnarray}
\label{coc2}
\sum_{\delta_1 \delta_2^{-1} =u } M_{\delta_1}^\dagger M_{\delta_2}
=\delta_{\{u=e\}} \mathbbm 1\\
\label{coc3}
\sum_{\delta_1 \delta_2^{-1} =u } M_{\delta_1} M_{\delta_2}^\dagger
=\delta_{\{u=e\}} \mathbbm 1
\end{eqnarray}
where both sums run over all pairs of elements in $\Delta$,
$u$ is any element in the set
\begin{eqnarray}
\Delta_2=\{ \delta \delta'^{-1}; \delta, \delta' \in \Delta\}
\end{eqnarray}
and $e$ is the neutral element in $\Gamma$.
The number of equations is twice the cardinality of $|\Delta_2|$ and
the number of terms in at least some of these equations will be larger than
one as soon as there exists closed paths of length 4 on the graph with an
alternating orientation, which in terms of the generators is
\begin{eqnarray}
\label{altp4}
\delta_1 \delta_2^{-1} \delta_4 \delta_3^{-1} =e
\end{eqnarray}
Because of this relation it will be sometimes useful to define the group
$\Gamma$ itself in terms of the ``free presentation''  
\begin{eqnarray}
\Gamma = \langle \Delta'| R\rangle 
\end{eqnarray}
where $\Delta'$ is a set of generators of a free group and $R$ is
the set (which may also be empty) of relations between the elements of
$\Delta'$ and their inverses which defines the structure of the group. To
define the Cayley graph \eref{xcg} and \eref{ecg} in the following we will use
the generating set $\Delta$  defined by
\begin{equation}\label{deltadef}
\Delta=\{\gamma:\gamma \in \Delta' \lor \gamma^{-1} \in \Delta'\}
\end{equation}
where $\Delta'$ is the generating set used in the free presentation of the
group. In particular $\Delta$ may contain at the same time a generator
and its inverse.\\
We now list some generic cases of Cayley groups.
\subsection{Cayley graphs of free groups}
As its name suggests, a free group is a group generated with a (finite) number
of generators with no relations between them
\begin{equation}\label{fg}
\Gamma =\langle \Delta' |- \rangle
\end{equation}
Lets consider the Cayley graph $C_{\Delta}(\Gamma)$ of the precedent group
defined by \eref{xcg}, \eref{ecg} and \eref{deltadef}. The two sets of
equations \eref{coc2} and \eref{coc3} can be written as:   
\begin{eqnarray}
\label{cocf1}
M_{\delta_1}^\dagger M_{\delta_2} &=& M_{\delta_1} M_{\delta_2}^\dagger = 0
\hbox{ for all } \delta_1 \ne \delta_2\\
\label{cocf2}
\sum_{\delta\in \Delta} M_{\delta} M_{\delta}^\dagger &=&
\sum_{\delta\in \Delta} M_{\delta}^\dagger M_{\delta} = \mathbbm 1
\end{eqnarray}
\begin{theo}
On the Cayley graph of the free group \eref{fg}, defined by \eref{xcg},
\eref{ecg} and \eref{deltadef}, the quantum walk evolution
operator \eref{eo} is unitary if and only if the internal operators are of the
form, 
\begin{equation}
\label{solcf}
M_{\delta} = U\, P_\delta
\end{equation}
where $U$ is a unitary matrix of dimension $dim(\mathcal H_I)$ and
$\{P_\delta\}_{\delta\in\Delta}$ is a complete family of orthogonal projectors,
\begin{equation}
\sum_{\delta\in\Delta} P_\delta = \mathbbm 1
\end{equation}
The internal space is of dimension larger or equal to $|\Delta|$.\\
\end{theo}
{\bf  Proof:} First, it is easy to see that \eref{solcf} is a
solution for \eref{cocf1}-\eref{cocf2}. Now suppose
\eref{cocf1}-\eref{cocf2}, these equations imply the following relation
between the images of the maps
\begin{eqnarray}
\label{cocf3}
\mathcal H_I = \oplus_{\delta\in\Delta} \mathcal Im(M_\delta) \\
\label{cocf4}
\mathcal H_I = \oplus_{\delta\in\Delta}  \mathcal Im(M_\delta^\dagger)
\end{eqnarray}
The fact that a direct sum appears in the right hand sides of
\eref{cocf3}-\eref{cocf4} is just a consequence of equations
\eref{cocf1} which make all subspaces pairwise orthogonal. The equality
(rather than an inclusion) is due to \eref{cocf2}.
Define $U\equiv \sum_\delta M_\delta$, an unitary matrix
by \eref{cocf1}-\eref{cocf2}, and $P_\delta$ as the orthogonal
projector on $\mathcal Im(M_\delta^\dagger)$, \eref{solcf} follows by
considering the elements of a vector basis compatible with the decomposition
\eref{cocf4} . The claim that \eref{solcf} is the general solution is thus
proven. $\square$\\ 
One should note however that the right hand side of \eref{solcf} could be
written in many other ways, for instance with its factors written in the
opposite order (which makes $P_\delta$ the projector on $Im(M_\delta)$).
When the rank of all matrices $M_{\delta}$ is fixed to 1, the dimension on  the
local Hilbert space takes its minimal value $dim(\mathcal H_I)= |\Delta|$, and
if a symmetric presentation for the group is chosen
(i.e: $\delta\in\Delta$ implies $\delta^{-1}\in\Delta$), the
standard definition of quantum ``coin'' solution  \cite{aakv} is
recovered. Besides these solutions, the only other possibility in the case of
free groups
consists in taking matrices $M_{\delta}$ of rank different from one and
possibly varying with $\delta$.\\
The case when the generating set that defines the Cayley graph contains the
group identity $e$ and at the same time some generators and their inverses is
slightly more involved because the group identity $e$ commutes with all the
elements in the group. If both a generator $\delta$ and his inverse
$\delta^{-1}$ are in $\Delta$ in addition to equations \eref{cocf1} one has
\begin{eqnarray}
\label{identity1}
&&M_{\delta}^\dagger M_{e}+M_{e}^\dagger M_{\delta^{-1}}=0\\
\label{identity2}
&& M_{e} M_{\delta}^\dagger + M_{\delta^{-1}} M_{e}^\dagger =0
\end{eqnarray}
for all $\delta\ne e$. Summing all equations in
\eref{identity1}, one gets $M_{e}^\dagger S =- S ^\dagger M_{e}$ where
$S=\sum_\delta M_\delta$.  Adding again two instances of equations
\eref{identity1} for both a given $\delta$ and its inverse $\delta^{-1}$
gives
\begin{eqnarray}
( M_{e}^\dagger S ) (P_\delta + P_{\delta^{-1}}) =
  (P_\delta + P_{\delta^{-1}}) ( M_{e}^\dagger S )
\end{eqnarray}
for all $\delta \not = e$. Thus $M_e^\dagger S$ is block diagonal in the
representation where all the orthogonal projectors $P_\delta$'s are
simultaneously diagonal. The problem can essentially be reduced to
the one dimensional case which we explore below. This is the first instance of
a solution to equations \eref{identity1} and \eref{identity2} different to the
solution \eref{solcf}, in the case when there is more than one non-zero term.
\subsubsection{One dimensional walks}
The simplest example is a quantum walk in one dimension. Lets consider the
group generated by one element $\Gamma=<\delta|->$ and the Cayley graph
obtained \eref{xcg}-\eref{ecg} using $\Gamma$ and the set
$\Delta=\{\delta,\delta^{-1}\}$. The minimal dimension of the internal
space is $2$ by the preceeding theorem and the form of the solution
follows equation \eref{solcf}. The evolution operator defined in
\eref{wcg} reads in this case   
\begin{eqnarray}
\label{wcf1d}
W = (U \otimes Id) (P_\delta \otimes T_\delta
+ P_{\delta^{-1}} \otimes T_{\delta^{-1}} )
\end{eqnarray}
where $U$ is a $2\times 2$ unitary matrix.
Two quantum walk evolution operators $W$ and $W'$ differing by an unitary
transformation $V$ would be equivalent, since this amounts
to a change of basis for the initial and final state. 
We will suppose $V$ of the form of a tensor product $A\otimes
\mathbbm{1}$.  Thus equation \eref{wcf1d} defines a family of inequivalent quantum walks
indexed by 4 real parameters: the 4 parameters associated with the unitary
matrix $U$ while the projectors $P_{\delta},P_{\delta^{-1}}$ become the
projectors over the spaces spanned by each one of the basis vectors.\\
A quantum walk can also be left-right symmetric if it is invariant, up to an
unitary transformation $S \otimes \mathbbm 1$, under the transformation
$T_\delta \leftrightarrow T_{\delta^{-1}}$. The family of inequivalent and
left-right symmetric quantum walks are of the reduced form described before
with $U$  
\begin{equation}
U=e^{i \delta}\left( \begin{array}{cc} \cos\frac{\theta}{2}& e^{i \alpha}
    \sin\frac{\theta}{2}\\-e^{-i\alpha} \sin\frac{\theta}{2}&
    \cos\frac{\theta}{2} \end{array} \right) 
\end{equation}
This defines then a 3 parameter family of inequivalent left-right symmetric
quantum walks. The unitary $S$ depends also on the 3 parameters.
When the identity appears in $\Delta$, two kinds of solutions
can be devised depending on whether the two terms appearing in
\eref{identity1}-\eref{identity2} are separately zero or not. In the first
case, one needs to add (at least) one state associated to the identity and the
evolution operator becomes 
\begin{eqnarray}
\label{wcf1de1}
W = (U \otimes Id) (P_\delta \otimes T_\delta
+ P_{\delta^{-1}} \otimes T_{\delta^{-1}} + P_e \otimes Id)
\end{eqnarray}
where $U$ is a $3\times 3$ unitary matrix, and appears just as a
simple extension of the previous example.
However, solutions exist with a two dimensional local Hilbert space,
and in such cases the evolution operator is
\begin{eqnarray}
\label{wcf1de2}
W = (U \otimes Id) \bigl( \cos(\theta) (P_\delta \otimes T_\delta
+ P_{\delta^{-1}} \otimes T_{\delta^{-1}}) + \sin(\theta) R_{\frac{\pi}{2}}
\otimes Id\bigr)
\end{eqnarray}
where $U$ is a $2\times 2$ unitary matrix, and $R_{\frac{\pi}{2}} = \left(
  \begin{array}{cc} 0 & 1 \\ -1 & 0 \end{array} \right)$. Equivalent solutions
with a two dimensional local Hilbert space were presented first in
\cite{meyer}. In conclusion, we also note that solution \eref{solcf} remains
valid when adding relations between generators. Thus such solutions exist for
all groups, in particular for free products of cyclic groups, 
\begin{equation*}
\Gamma=\langle \delta_1, \cdots , \delta_l | \delta_1^{q_1} = \cdots
\delta_l^{q_l} = e \rangle
\end{equation*}
and for free Abelian groups,
\begin{equation*}
\Gamma=\langle \delta_1, \cdots ,\delta_l  | \delta_i \, \delta_j \,
\delta_i^{-1} \delta_j^{-1} = e \;\forall i,j \in \{1,\cdots,l\}\rangle
\end{equation*}
which we consider in the next section.
\subsection{Cayley graphs of free Abelian groups}
One should note that the commutation relations between
elements from the set of generators and their inverses, for instance
\begin{eqnarray}
\delta_1 \delta_2^{-1}=\delta_2^{-1}\delta_1
\end{eqnarray}
do not necessarily imply the existence of a closed path on the graph
with alternate orientation of the edges \eref{altp4}, except in the case when
the inverses of the elements of $\Delta$ are themselves in $\Delta$. The group is defined by:
\begin{equation}\label{sag1}
\Gamma =\langle  \delta_1, \dots ,\delta_n |
\delta_i\delta_j\delta_i^{-1}\delta_j^{-1}=e \,\forall i,j \in \{1, \dots ,n\}
\rangle
\end{equation}
and the set used to construct the Cayley graph is
\begin{equation}\label{sag2}
\Delta=\{ \delta_1, \dots ,\delta_n,\delta_1^{-1}, \dots ,\delta_n^{-1} \}
\end{equation}
In such a case, equations \eref{coc2}-\eref{coc3} read
\begin{eqnarray}
\label{cocfc1}
M_{\delta_i}^\dagger M_{\delta_j} + M_{\delta_j^{-1}}^\dagger M_{\delta_i^{-1}}
&=& 0 \hbox{ for all } \delta_i \ne \delta_j\\
\label{cocfc2}
M_{\delta_i} M_{\delta_j}^\dagger +M_{\delta_j^{-1}} M_{\delta_i^{-1}}^\dagger
&=& 0 \hbox{ for all } \delta_i \ne \delta_j\\
\label{cocfc3}
\sum_{\delta\in \Delta} M_{\delta} M_{\delta}^\dagger =
\sum_{\delta\in \Delta} M_{\delta}^\dagger M_{\delta} &=& \mathbbm 1
\end{eqnarray}
When $\delta_j= \delta_i^{-1}$, equations \eref{cocfc1}-\eref{cocfc2}
contain a single term and read
\begin{eqnarray}
\label{cocfc4}
M_{\delta_i}^\dagger M_{\delta_i^{-1}} = M_{\delta_i^{-1}} M_{\delta_i}^\dagger &=& 0
\end{eqnarray}
These are much less restrictive conditions than \eref{cocf1}-\eref{cocf2}, and
we lack here the decomposition of $\mathcal H_I$ into orthogonal subspaces
which allowed us to give a general answer in the case of free groups.
We only notice that equations \eref{cocfc1}-\eref{cocfc2} are equivalent
to the following
\begin{eqnarray}
\bigl(\sum_{\delta\in A} \lambda_\delta M_\delta^\dagger) \bigr)
\bigl(\sum_{\delta\in A} \lambda_\delta M_\delta^{-1}) \bigr)=
\bigl(\sum_{\delta\in A} \lambda_\delta M_\delta^{-1}) \bigr)
\bigl(\sum_{\delta\in A} \lambda_\delta M_\delta^\dagger) \bigr) = 0
\end{eqnarray}
for all subset $A\in \Delta$ such that $\delta\in A \Rightarrow
\delta^{-1}\not\in A$ and for all families of real parameters
$\{\lambda_\delta\}_{\delta\in A}$.\\
Equations \eref{cocfc1}-\eref{cocfc2} imply
the following proposition which will help us classify the solutions.
\begin{prop}
Let $C_{\Delta}(\Gamma)$ be the Cayley graph of the free abelian group with
$n$ generators (\eref{sag1}-\eref{sag2}) . If a quantum walk operator \eref{eo} defined on $G$ is
unitary then the image subspaces of any two internal operators $M_{\delta_i}$
and $M_{\delta_j}$ are either orthogonal or contain a common vector
subspace. The same implication is valid for the image subspace of their
conjugates $M_{\delta_i}^\dagger$ and $M_{\delta_j}^\dagger$: 
\begin{eqnarray}
\label{ortho}
\bigl(\mathcal Im(M_{\delta_i}) \cap \mathcal Im(M_{\delta_j})
=\{0\}\bigr)\Rightarrow 
\mathcal Im(M_{\delta_i}) \bot \,\mathcal Im(M_{\delta_j})\\
\label{orthodag}
\bigl(\mathcal Im(M_{\delta_i}^\dagger) \cap \mathcal Im(M_{\delta_j}^\dagger)
=\{0\}\bigr)\Rightarrow 
\mathcal Im(M_{\delta_i}^\dagger) \bot \,\mathcal Im(M_{\delta_j}^\dagger)
\end{eqnarray}
\end{prop}
{\bf Proof:}
Using \eref{cocfc2} for a pair $\delta_i$,
$\delta_j^{-1}$, one has
\begin{eqnarray}
\label{intersection}
\mathcal Im(M_{\delta_i} M_{\delta_j^{-1}}^\dagger) =
\mathcal Im(M_{\delta_j} M_{\delta_i^{-1}}^\dagger)
\end{eqnarray}
and thus
\begin{eqnarray}
\mathcal Im(M_{\delta_i} M_{\delta_j^{-1}}^\dagger)
\subset
\bigl(\mathcal Im(M_{\delta_i}) \cap \mathcal Im(M_{\delta_j})\bigr)
\end{eqnarray}
Suppose now that $\mathcal Im(M_{\delta_i})$ and $ \mathcal Im(M_{\delta_j})$
have no common vector subspace. Thus
$M_{\delta_i} M_{\delta_j^{-1}}^\dagger = 0$, which can be written
\begin{eqnarray}
\mathcal Im(M_{\delta_i}^\dagger) \bot
\mathcal Im(M_{\delta_j^{-1}}^\dagger)
\end{eqnarray}
and more particularly
\begin{eqnarray}
\mathcal Im(M_{\delta_i}^\dagger M_{\delta_j}) \bot
\mathcal Im(M_{\delta_j^{-1}}^\dagger M_{\delta_i^{-1}})
\end{eqnarray}
Since the two subspaces are equal (by \eref{cocfc4}) and orthogonal,
they are equal to the null vector space and hence we have again
$M_{\delta_i}^\dagger M_{\delta_j} =0$, and finally
\begin{eqnarray}
\mathcal Im(M_{\delta_i}) \bot \mathcal Im(M_{\delta_j})
\end{eqnarray}
The implication \eref{ortho} is thus proven. The proof of \eref{orthodag}
is equivalent, beginning with equation \eref{cocfc1} instead of
\eref{cocfc2}. $\square$ \\
We can use Proposition (1) to find solutions with an internal space dimension
smaller than the number of generators in the following way.
First we can write
\begin{eqnarray}
\label{dimcc}
{\rm dim}(\mathcal H_I) \ge \sup_{\delta_i,\delta_j}\bigl\{
\sum_{\epsilon_1,\epsilon_2=\pm 1}
{\rm dim}\bigl(\mathcal Im(M_{\delta_i^{\epsilon_1}}) \cap
\mathcal Im(M_{\delta_j^{\epsilon_2}}) \bigr)\bigr\}
\end{eqnarray}
where the $\sup$ runs over all pairs $\delta_i, \delta_j$ such that both
$\delta_i \ne\delta_j$ and $\delta_i \ne\delta_j^{-1}$. This inequality
is true since the four sets appearing in the right hand side are pairwise
orthogonal by \eref{cocfc4}.
A similar equation could be written involving the $M^\dagger$'s.
Suppose now that the supremum on the right hand side of \eref{dimcc} is zero,
hence giving no direct condition on the dimension of $dim(\mathcal H_I)$.
In such a case, all vector subspaces are orthogonal by \eref{ortho},
which imply $dim(\mathcal H_I) \ge \vert \Delta\vert $. Hence, a necessary condition for the existence of
quantum walks with a smaller internal space is that some of the
intersections in the sum \eref{dimcc} are non empty.
In the following we give some examples.

\subsubsection{A two dimensional walk with a two dimensional internal space.}
We consider here the group $\Gamma=\langle \delta_1,\delta_2  | \delta_1
\delta_2 \delta_1^{-1} \delta_2^{-1} = e \rangle$, a symmetric set
$\Delta=\{\delta_1,\delta_1^{-1}, \delta_2, \delta_2^{-1}\}$  and define a
quantum walk over the associated Cayley graph through the evolution operator
\eref{wcg} which reads here
\begin{eqnarray}
W=M_{\delta_1} \otimes T_{\delta_1}+ M_{\delta_1^{-1}} \otimes
T_{\delta_1^{-1}}+M_{\delta_2} \otimes T_{\delta_2} + M_{\delta_2^{-1}} \otimes T_{\delta_2^{-1}}
\end{eqnarray}
We suppose that the rank of each matrix $M_{\delta_i}$ is one. In order to impose $dim(\mathcal H_I) =2$,
we require that at least two terms in the right hand side of \eref{dimcc} are zero for each possible pair
of generators $\delta_i,\delta_j$.
We obtain two solutions which transform one derived from the other by changing
$\delta_1$ and $\delta_1^{-1}$. Up to an unitary transformation,
the solution is
\begin{eqnarray*}
\fl M_{\delta_1}=UP_1VP_1 \quad M_{\delta_1^{-1}}=UP_2VP_2 \quad M_{\delta_2}=UP_1VP_2 \quad
M_{\delta_2^{-1}} =UP_2VP_1
\end{eqnarray*}
where $U$ and $V$ are two unitary matrices and $P_1$,$P_2$ two orthogonal
projectors. The evolution operator factorises into a product of two
one-dimensional operators 
\begin{eqnarray*}
\fl W= ( U\otimes 1) ( P_1 \otimes (T_{\delta_1}T_{\delta_2})^{\frac{1}{2}} +
 P_2 \otimes (T_{\delta_1^{-1}}T_{\delta_2^{-1}})^{\frac{1}{2}})\\
 (V\otimes 1) (P_1 \otimes (T_{\delta_1}T_{\delta_2^{-1}})^{\frac{1}{2}}
 +P_2 \otimes (T_{\delta_1^{-1}}T_{\delta_2})^{\frac{1}{2}})
\end{eqnarray*}
However a quantum walk with a two dimensional
internal space which is symmetric by inversion of only one of the axes or by a
rotation of angle $\frac{\pi}{2}$ does not exist.\\
This solution generalises in higher dimensions: \\
\begin{prop} Let $C_{\Delta}(\Gamma)$ be the Cayley graph of the free abelian
  group with $n$ generators and symmetric presentation \eref{sag1} and
  \eref{sag2}. Then there exists a unitary quantum walk operator \eref{eo} on
  $G$ such that the dimension of the internal space is $n$ if $n$ is even and
  $n+1$ if $n$ is odd. 
\end{prop}
{\bf Proof:}
Suppose $n$ even. We consider an internal space of dimension $n$ and decompose
it as a direct sum of two dimensional subspaces. We associate to each of
these subspaces one different pair of generators. For such a pair
$(\delta_i,\delta_j)$, the four operators
$M_{\delta_i},M_{\delta_j},M_{\delta_i^{-1}},M_{\delta_j^{-1}}$ act non
trivially only on the associated two-dimensional subspace
and can be constructed in the same way as the internal operators of the
previous example of two dimensional walk. The dimension of the internal space for such a quantum walk is then half the
dimension of the free form solution. Suppose now $n$ odd. We can repeat the previous construction for $n-1$
generators, and add a two dimensional space where the internal operators
associated to the last generator will have the form of the internal operators
of a one dimensional walk. All the internal operators then verify the condition
equations \eref{cocfc1}-\eref{cocfc3}. $\square$\\
\subsubsection{Two dimensional walks with a four dimensional internal space.}
The impossibility of having a fully symmetric quantum walk does not hold when
taking a four dimensional internal space. One possibility is to  suppose that
all the intersections involved in \eref{dimcc} are of dimension zero, in this
case $dim(\mathcal H_I) \ge \vert \Delta \vert = 4$ and the minimal choice
of the dimension leads to an evolution operator $W=\sum_\delta P_\delta U
\otimes T_\delta$ where U is a four dimensional unitary matrix. 
The other possibility is to  suppose that all the intersections
involved in \eref{dimcc} are of dimension one. In this case
the minimal dimension of the internal space is also four. A
simple choice of matrices of rank two verifying all the conditions
\eref{cocfc1}-\eref{cocfc3} is:
\begin{eqnarray}
M_{\delta_1} &=&  \frac{1}{\sqrt{2} }(\,|u_1\rangle \langle v_1 | + |u_2\rangle \langle v_3 |\,)\\
M_{\delta_1^{-1}} &=&  \frac{1}{\sqrt{2} }(\,- |u_3\rangle \langle v_4 | + |u_4\rangle \langle v_2 |\,)\\
M_{\delta_2} &=&  \frac{1}{\sqrt{2} }(\, |u_1\rangle \langle v_2 | +  |u_3\rangle \langle v_3 |\,) \\
M_{\delta_2^{-1}} &=&  \frac{1}{\sqrt{2} }(\,- |u_4\rangle \langle v_1 | + |u_2\rangle \langle v_4 |\,)
\end{eqnarray}
where  $\{ |u_i\rangle\}_{i=1,\cdot,4}$ and $\{|v_i\rangle\}_{i=1,\cdot,4}$
are two orthonormal bases of  $\mathcal H_I$.
In the following we give the explicit form of the evolution
operator supposing that the rank of the matrices $M_\delta$ is one
and that the walk is symmetric.
A permutation of the
 vertex set $\Pi$ is associated with a spatial transformation. As in the one dimensional case, the walk is symmetric under
 this transformation if there exists an unitary $S$ such that $(S \otimes
 \Pi)^{\dagger} W (S \otimes \Pi) = W$. In other words, if the initial
 condition is modified by the transformation $S \otimes \Pi$, the
wave function at any time can be deduced from the unmodified wave
function by application of the same transformation.
We impose invariance under the symmetries of the square lattice by considering
the two transformations, $S_i \otimes \Pi_i$ and $S_r \otimes \Pi_r$, being
respectively the representation of the inversion along the $x$ axis
and the rotation by ${\frac{\pi}{2}}$. The symmetry condition makes $U$ reduce
to a product $U=D^{-1} U_0 D$ where $D$ is a diagonal unitary matrix depending
on four real parameters and $U_0$ takes the form:
 \begin{displaymath}
 U_0 = \left( \begin{array}{cccc} a & b & c & c \\ b & a & c & c \\ c &
 c & a  & b \\  c & c &  b  & a
 \end{array} \right) \end{displaymath}
The matrix $U_0$ depends on $3$ parameters by the unitarity
condition. The matrices $S_1$ and $S_2$ depend on the same parameters as the
matrix $D$. Then choosing these four parameters equal to one
reduces the walk  operator to  $W=\sum_i P_i U_0 \otimes Ti$ and the
matrices $S_1$ and $S_2$ are just the inverse permutation of the
generators associated to the spatial transformation.
\subsubsection{A three dimensional walk with a four dimensional internal
  space.} It has been shown that no nontrivial solution exists in three
dimensions with a two dimensional internal space\cite{Bialynicki}. 
In the following we give solutions on $\mathbb Z^3$ with a
four dimensional internal space.
The starting point is again equation \eref{dimcc}. Taking matrices of rank two would
not break this condition for $dim(\mathcal H_I)$ provided that each term on
the left hand side of \eref{dimcc} is one. Here we thus give the general solution
for rank two matrices. Let $\Delta=\{\delta_1,\delta_2,\delta_3,\delta_1^{-1},\delta_2^{-1},\delta_3^{-1}\}$.
Defines two orthonormal bases $\{ |u_i\rangle\}_{i=1,\cdot,4}$ and $\{ |v_i\rangle\}_{i=1,\cdot,4}$.
Now construct six matrices of rank 2 indexed by the elements of $\Delta$ in the form
\begin{eqnarray}
M_{\delta_1} &=& \alpha_1 |u_1\rangle \langle v_2 | + \beta_1 |u_2\rangle \langle v_1 |\\
M_{\delta_1^{-1}} &=& \gamma_1 |u_3\rangle \langle v_4 | + \delta_1 |u_4\rangle \langle v_3 |\\
M_{\delta_2} &=& \alpha_2 |u_1\rangle \langle v_3 | + \gamma_2 |u_3\rangle \langle v_1 | \\
M_{\delta_2^{-1}} &=& \beta_2 |u_2\rangle \langle v_4 | + \delta_2 |u_4\rangle \langle v_2 |\\
M_{\delta_3} &=& \alpha_3 |u_1\rangle \langle v_4 | + \delta_3 |u_4\rangle \langle v_1 |\\
M_{\delta_3^{-1}} &=& \beta_3 |u_2\rangle \langle v_3 | + \gamma_3 |u_3\rangle \langle v_2 |
\end{eqnarray}
It is clear that such a choice solves equations \eref{cocfc4}. The other
equations are solved by taking
\begin{eqnarray}
\alpha_2 =\lambda \alpha_1\qquad; \qquad \alpha_3= \mu \alpha_1\\
\beta_2 =\bar\lambda\nu \beta_1\qquad; \qquad \beta_3= -\bar\mu \nu \beta_1\\
\gamma_2 =-\lambda\bar\nu \gamma_1\qquad; \qquad \gamma_3= -\bar\mu \gamma_1\\
\delta_2 =-\bar\lambda \delta_1\qquad; \qquad \delta_3= \mu\bar\nu \delta_1
\end{eqnarray}

where $|\nu|^2=1$, $\lambda, \mu \in \mathbb C$ and
\begin{eqnarray}
|\alpha_1|=|\beta_1|=|\gamma_1|=|\delta_1|=\frac{1}{\sqrt{1+|\lambda|^2+|\mu|^2}}
\end{eqnarray}
\subsection{Cayley graphs with multiply connected second neighbours}
In this section we consider Cayley graphs in which any second neighbour
is connected by at least two alternating paths. They might be of interest
since the condition equations contain at least two terms. Here,
we only consider two examples in which each second neighbour is connected by
at least two alternate paths. Both are interesting in their own right:
the first one admits a scalar solution, while the other admits solutions in
terms of a Clifford algebra.
\subsubsection{A simple one dimensional example}
Let us consider the commutative group with two generators \eref{sag1} with one
more relation $\delta_1^2  = \delta_2^2$ in the presentation and as
defining set for the Cayley graph $\Delta= \{ \delta_1,
\delta_2,\delta_1^{-1},\delta_2^{-1} \}$.\\
The four matrices $M_\delta$ have to be taken as solutions of the
four equations:
\begin{eqnarray}
&&M_{\delta_1}^\dagger M_{\delta_1^{-1}}
 +  M_{\delta_2}^\dagger M_{\delta_2^{-1}} =
 M_{\delta_1^{-1}} M_{\delta_1}^\dagger +
 M_{\delta_2^{-1}} M_{\delta_2}^\dagger  = 0\\
&& M_{\delta_1}^\dagger M_{\delta_2^{-1}}
 +  M_{\delta_2}^\dagger M_{\delta_1^{-1}} =
 M_{\delta_2^{-1}} M_{\delta_1}^\dagger
 + M_{\delta_1^{-1}} M_{\delta_2}^\dagger = 0\\
&&M_{\delta_1}^\dagger M_{\delta_2}
+  M_{\delta_2}^\dagger M_{\delta_1}
 + M_{\delta_1^{-1}}^\dagger M_{\delta_2^{-1}}
 +  M_{\delta_2^{-1}}^\dagger M_{\delta_1^{-1}} =0\\
&&M_{\delta_2} M_{\delta_1}^\dagger
+ M_{\delta_1} M_{\delta_2}^\dagger
+ M_{\delta_2^{-1}} M_{\delta_1^{-1}}^\dagger
+ M_{\delta_1^{-1}} M_{\delta_2^{-1}}^\dagger =0\\
&&\sum_\delta M_\delta^\dagger  M_\delta =
 \sum_\delta M_\delta M_\delta^\dagger=\mathbbm 1
\end{eqnarray}
This set of equations admits solutions with a one dimensional internal space,
and the evolution operator can be written as
\begin{eqnarray}
W= \frac{1}{2} \bigl(e^{i\theta} (\tau_1\pm\tau_2) +
e^{i\varphi} (\tau_1^{-1}\mp\tau_2^{-1}) \bigr)
\end{eqnarray}
where $\tau_1$ and $\tau_2$ are the displacements in the directions
$\delta_1$ and $\delta_2$.
However, as can be seen from the form of the evolution operator, this
example is equivalent to a quantum walk on $\mathbb Z$ with a two
dimensional internal space by grouping together pairs of second neighbours.
What is interesting here is that even on a graph where all sites
are equivalent, there may exist scalar solutions provided all second
neighbours are multiply connected. The minimal
dimension of the internal space would still however have to be questioned
since it strongly depends on the choice of the graph and various descriptions
appear to be equivalent.
\subsubsection{The hypercube}
We consider the group presentation
\begin{equation}\label{hyp}
\Gamma=\langle \delta_1, \cdots \delta_n |\delta_i^2 = e \,\forall i;\,
\delta_i\delta_j\delta_i^{-1}\delta_j^{-1} = e \,\forall i\ne j\rangle
\end{equation}
whose Cayley graph is the hypercube in $n$ dimensions. The condition equations
become: 
\begin{eqnarray}
M_{\delta_i}^\dagger M_{\delta_j}+M_{\delta_j}^\dagger M_{\delta_i}=0 \label{hyp1}\\
M_{\delta_i} M_{\delta_j}^\dagger+M_{\delta_j} M_{\delta_i}^\dagger=0\label{hyp2}\\
\sum_\delta M_\delta^\dagger M_\delta =\mathbbm{1}
\end{eqnarray}
Equations \eref{hyp1} and \eref{hyp2} are valid for all pairs of generators
$\delta_i,\delta_j$. Solutions originating from those for a
free group of n generators have been studied by various authors (\cite{kempehyperc}-\cite{Moore}).
\begin{prop}
There exists a unitary quantum walk operator \eref{eo} on the Cayley graph of
the group \eref{hyp} such that the internal operators are of the form
$M_{\delta_i}= \frac{1}{\sqrt n} \sigma_i U$ where $U$ is a unitary matrix of dimension $dim(\mathcal H_I)$ and $\{\sigma_1 \dots \sigma_n\}$ is a set of anticommuting matrices. 
\end{prop}
{\bf Proof:} If one requires that all the matrices $M_\delta$ be Hermitian (or
anti-hermitian) then the first set of equations \eref{hyp1}-\eref{hyp2} takes
the form of an anticommutation relation between all pairs of
matrices. Hermitian anticommuting matrices generate a Clifford algebra, it is
therefore natural to find solutions among their matrix 
representations. Let $\{\sigma_1 \dots \sigma_n\}$ such a set of anticommuting
matrices and $U$ an unitary matrix. A possible choice for the matrices $M_\delta$ is then $M_{\delta_i}= \frac{1}{\sqrt n} \sigma_i U$. $\square$ \\
For example, equations for $n=3$ are solved by
$M_i=\frac{1}{\sqrt 3} \sigma_i U$ where each $\sigma_i$ is one of the
three Pauli matrices and U a unitary matrix in two dimensions.
While the dimension of the matrix representation is rather large,
(at least $2^{\lbrack\frac{n}{2}\rbrack}$), such solution may nevertheless
be useful. 
\section{A generalised model of quantum walk}
A quantum walk is a model for the motion of a quantum particle jumping (quantically) over a graph.
A particle having a fixed number of internal degrees of freedom, one is naturally led to attach
to each point $x$ of the graph a copy of some Hilbert space $\mathcal H_I$ describing them.
This is obviously not a necessary hypothesis in the context of a network of quantum processors,
and even if we will retain here most of the terminology of quantum walks, we will not base our approach in this section on the interpretation of our
quantum object as a physical particle. A second important property is the
choice of a discrete time evolution, again motivated by the idea that
quantum processors as their classical equivalents would exchange information
at discrete times.\\ 
We will continue to consider discrete time evolution but we
want to note that quantum walks with continuous time has also been introduced
in the context of quantum algorithmics \cite{FG98} \cite{CFG01}. As for the
discrete time model, the succes of these walks performing particular tasks
is dependent on characteristics such as the initial vector state \cite{BM04},
thus indicating that a classification of this model may also be of some
interest. Some properties of one dimensional walks have been determined as for
example the revival time \cite{BM05} and a limit theorem demostrated
by \cite{Kon05}.\\     
We consider an oriented graph $G=(X,E)$, $X$ the set of vertices, and $E$ the set of oriented edges. To each vertex $x\in X$,
we attach a (finite) Hilbert space $\mathcal H_x$, and define the quantum evolution over $\mathcal H = \oplus_{x \in X}\mathcal H_x$
as follows:
For each oriented pair $(x,y)$, we define a linear map $M_{x,y}$ from $\mathcal H_x$ to $\mathcal H_y$, extend it on
$\mathcal H$ by setting $M_{x,y}=0$ on $\mathcal H_x^\bot$. We define its conjugate $M_{x,y}^\dagger$ as
the map such that
\begin{eqnarray}
\langle \Psi' | M_{x,y}\Psi \rangle = \langle  M_{x,y}^\dagger \Psi'|\Psi \rangle
\end{eqnarray}
for all $|\Psi \rangle$, $|\Psi' \rangle$ in $\mathcal H$.
Then we define the evolution of the quantum walk over $\mathcal H$
as:
\begin{eqnarray}
|\Psi(t+1)\rangle = W |\Psi(t)\rangle
\end{eqnarray}
where $|\Psi(t)\rangle$
is the state of the system at time $t$ and $W$ is the unitary operator
\begin{eqnarray}
W =\sum_{(x, y)\in E} M_{x,y}
\end{eqnarray}
In order to restrict the sum to the pairs of neighbouring sites and
impose $W$ to be
unitary we require the following three properties:
\begin{eqnarray}
\label{cof1}
M_{x,y} \ne 0 \hbox{ if and only if } (x,y)\in E
\end{eqnarray}
\begin{eqnarray}
\label{cof2}
\sum_y  M_{x,y}^\dagger M_{z,y} = \sum_y  M_{y,x} M_{y,z} ^\dagger = 0 \hbox{ for all } x\ne z
\end{eqnarray}
\begin{eqnarray}
\label{cof3}
\sum_y  M_{x,y}^\dagger M_{x,y} = \sum_y  M_{y,x} M_{y,x} ^\dagger =  {\bf 1}_x
\end{eqnarray}
where ${\bf 1}_x$ is the projector over $\mathcal H_x$.
Conditions \eref{cof2}
and \eref{cof3} are necessary and sufficient conditions for $W$ to be
unitary.
Here, it is already interesting
to note that even in this more general context quantum ``coin'' solutions
exist provided that on each site the number of incoming edges equals the number
of outgoing ones. The construction can be done
in the following way:
we first set the dimension of all local Hilbert spaces equal to the number of
incoming (or equivalently outgoing ) neighbours,
\begin{eqnarray}
dim(\mathcal H_x) = |E_x^{in}| = |E_x^{out}|
\end{eqnarray}
where we have set
\begin{eqnarray}
E_x^{in} = \{y\in X : (y,x)\in E\}\\
E_x^{out} = \{y\in X : (x,y)\in E\}
\end{eqnarray}
For all $x\in X$ we fix two orthonormal basis $\mathcal B_x^{in}$ and $\mathcal B_x^{out}$ in $\mathcal H_x$
an label its elements using the list of neighbours,
\begin{eqnarray}
\mathcal B_x^{in} =\bigl\{ |\varphi_x^{in}(y)\rangle\bigr\}_{y\in E_x^{in}}
\end{eqnarray}
\begin{eqnarray}
\mathcal B_x^{out} =\bigl\{ |\varphi_x^{out}(y)\rangle\bigr\}_{y\in E_x^{out}}
\end{eqnarray}
Now setting
\begin{eqnarray}
M_{x,y} = |\varphi_y^{in}(x)\rangle \langle \varphi_x^{out}(y)|
\end{eqnarray}
just satisfies all conditions \eref{cof2}, \eref{cof3} and defines a general quantum ``coin'' solution
even outside the context of a quantum particle on a lattice. In fact we get some more insight on
how such solutions works from the point of view of a quantum network:
first, each node splits the (partial) wave function
along the vectors of a fixed basis $\mathcal B_x^{out}$
and send the resulting complex number to each of its neighbours;
then a (partial) wave function is recomposed using the received numbers
and the other fixed basis $\mathcal B_x^{in}$.
We now want to recover previous definition of quantum walks on a Cayley graph,
so we naturally suppose that the properties of the graph are transferred to the
walk.
In particular all local Hilbert spaces are copies of the same space,
\begin{eqnarray}
\mathcal H_x = \mathcal H_0
\end{eqnarray}
for all $x$ in $X$ and the complete Hilbert space is equivalent to the direct
product of the local space $\mathcal H_0$ with a position space $\mathcal
H_X$.
\begin{eqnarray}
\mathcal H \approx \mathcal H_0 \otimes \mathcal H_X
\end{eqnarray}
Furthermore, the maps $M_{x,y}$ will depend only on the
edge colour and direction of the edge $(x,y)$ (i.e. only on the generator $\delta= x^{-1} y$)
and not in the starting vertex $x$:
\begin{eqnarray}
M_{x,y}= T_{0,y} M_{x^{-1} y} T_{x,0}\hbox{ for all } (x,y) \in E
\end{eqnarray}
where $M_{x^{-1} y}$ is a map on $\mathcal H_0$ and $T_{x,y}$ is the canonical
shift map sending $\mathcal H_x$ onto $\mathcal H_y$. Thus the evolution
operator $W$ on $\mathcal H$ as a product space reads
\begin{eqnarray}
W = \sum_{\delta\in\Delta} M_\delta \otimes T_\delta
\end{eqnarray}

\section{Conclusion}
We have considered quantum walks on Cayley graphs of groups and addressed the
problem of classifying them as a function of the group presentation
and the choice of the internal space. A first result is that
the smallest possible dimension of the internal space depends strongly on the
generating set chosen for constructing the Cayley graph.
In the case of free groups, we succeeded in classifying all possible solutions.
Standard quantum walk definition is recovered and correspond to an internal
space of dimension equal to the number of neighbours (its smallest value) and
a free group with a set of generators containing elements of the group
different from the identity. 
When the identity element is present in the generating set used to define the
Cayley graph of a free grop, or on other
Cayley graphs, we showed that different solutions do exist
for which we give a partial characterisation. 
We presented a few examples of solutions which does not enter in the
previously known solutions and which become available as soon as there exist
closed paths of length 4 on the graph, with alternating orientation.
In particular, we found solutions with a smaller internal dimension that
what is usually expected and a new kind of quantum walks on the hypercube
based on Clifford algebra representation.
We hope that these new possibilities will prove useful in the
context of the relationship between quantum walks and quantum algorithms.

\section{Acknowledgements}
We are grateful to Z. Nagy and F. Millet for helping us in finding
Clifford solutions on the hypercube and to J. Avan, J.-P. Kownacki,
M. Sch\"urmann and N. Weatherall for useful discussions.\\
Research partially supported by European Commission HPRN-CT-2002-00279, RTN QP
Applications.

\end{document}